\documentclass[10pt, english,prd,
superscriptaddress,altaffilletter,
nofootinbib,preprintnumbers,showpacs,floatfix,notitlepage,twocolumn]{revtex4-1}

\usepackage{amsmath}
\usepackage{float}
\usepackage{multirow}
\usepackage{comment}
\usepackage{overpic}
\usepackage{ifxetex}
\ifxetex
   \usepackage{polyglossia}
   \setdefaultlanguage{english}
   \usepackage{fontspec}
\else
   \usepackage[english]{babel}
   \usepackage[utf8]{inputenc}
\fi

\usepackage{cancel}

\usepackage{verbatim} 

\newcommand{\be}{\begin{equation}}
\newcommand{\ee}{\end{equation}}
\newcommand{\bea}{\begin{eqnarray}}
\newcommand{\eea}{\end{eqnarray}}

\DeclareMathOperator{\arctanh}{arctanh}

\usepackage{color}
\usepackage{ amssymb }
\usepackage[dvipsnames]{xcolor}
\begin{document}

\title{DGP and DGPish cosmologies from $f(Q)$ actions}

\author{Ismael Ayuso} 
\email{iayuso@fc.ul.pt}
\affiliation{Departamento de Física and Instituto de Astrofísica e Ciências do Espaço, Faculdade de Ciências, Universidade de Lisboa, Edifício C8, Campo Grande, 1769-016 Lisboa, Portugal}
\author{Ruth Lazkoz}
\email{ruth.lazkoz@ehu.es}
\affiliation{Department of Physics, University of the Basque Country UPV/EHU, P.O. Box 644, 48080 Bilbao, Spain}
\author{Jos\'e Pedro Mimoso}
\email{jpmimoso@fc.ul.pt}
\affiliation{Departamento de Física and Instituto de Astrofísica e Ciências do Espaço, Faculdade de Ciências, Universidade de Lisboa, Edifício C8, Campo Grande, 1769-016 Lisboa, Portugal}

\date{\today}

\begin{abstract}

In this work we explore and test new formulations   of  cosmological scenarios in $f(Q)$ theories. In these settings, the non-metricity scalar ($Q$) is the main source of gravity  and Friedmann equations are modified to account for the associated degrees of freedom. This work focuses first on the derivation, and then theoretical and observational analysis of two  such (new) exact cosmological models; they both display  a non-standard behaviour in which an additional parameter encoding non-metricity effects acts in the fashion of a screened cosmological constant. One of the new settings has the same background evolution as the well know DGP cosmological model, while the other resembles the former considerably, although its origin is purely phenomenological. We use the Markov Chain Montecarlo method combined with standard statistical techniques to perform observational astrophysical tests relying upon background data, specifically these are Type Ia Supernovae luminosities and direct Hubble data (from cosmic clocks), along with Cosmic Microwave Background shift and Baryon Acoustic Oscillations data. In addition, we compute some of the cosmographic parameters and other discriminators  with the purpose of refining our knowledge about these models in the light of their theoretical and observational signatures, and this allows for a better  comparison with the (concordance) $\Lambda$CDM setup. We conclude that these scenarios do not show signatures indicating a departure from the  $\Lambda$CDM behaviour.
\end{abstract}

\maketitle

\section{Introduction}

It is widely understood that in order to
build on a deeper comprehension of our universe we must first understand the force that dominates it, the gravitational interaction, which happens to be the weakest force in Nature. General Relativity (GR) \cite{Einstein:1915ca,Einstein:1915by} has stood triumphantly \cite{Einstein:1915ca,Einstein:1915by} as the former's best description, after being subject to  all  tests imaginable  up to the solar system scale \cite{Will:2014kxa, Berti:2015itd}.

Nonetheless, the  physics of the cosmos on larger scales poses unresolved questions demanding  to  consider  non-standard types of energy and matter acting as sources of geometry. These days the indisputable approach is one that  describes the background of our GR ruled large-scale universe as homogeneous and  isotropic. Under those assumptions we can rest upon the $\Lambda$CDM model to offer a very decent narrative of our Universe fuelled by those unusual components we have referred to before: dark matter and dark energy (see for instance \cite{Abdalla:2020ypg}). The first one offers a competent explanation for the formation and evolution of structures, while the second one does a very good job at explaining the current accelerated expansion of the Universe.

However, if we examine deeply the $\Lambda$CDM model, we realize that its two main components introduce new problems \cite{Carroll:2000fy, Peebles:2002gy, Weinberg:1988cp, Padmanabhan:2002ji, Carroll:2003qq, Bianchi:2010uw, Bull:2015stt}. There are several proposal to resolve it, in particular, analyzing and focusing on the specific type of dark energy \cite{Frampton:2004nh, Brax:2017idh}, which could end up being not so simple and at the same time not so ``puzzling" as a bare cosmological constant, the often also called $\Lambda$ term. But the spirit of this work is to try and reproduce  equivalent effects (and therefore the associated accelerated expansion) by adopting a different perspective. 

Our framework will be that of modified theories of gravity, in which GR and its assumptions are (slightly) tweaked  \cite{Capozziello:2002rd, Clifton:2011jh, Joyce:2014kja, Harko:2018ayt}. Among the immense zoo of possibilities,  we will embrace that of metric-affine geometry, which generalizes the Riemannian geometry approach adopted in GR. This relies upon letting the connection be a non-standard free variable at the same level as the metric in GR
(and forgetting the Levi-Civita connection, even though it is possible to recover it for some specific cases). Obviously, this freedom induces a richer phenomenology of the theories associated with the transformation of physico-mathematical objects along a displacement \cite{BeltranJimenez:2019tjy,Jimenez:2019ghw,Bahamonde:2021gfp}.

Specifically, in this work we will concentrate on modified theories built from the non-metricity tensor $Q_{\alpha\mu\nu}\equiv\nabla_\alpha g_{\mu\nu}$, and more specifically from the non-metricity scalar \cite{Mol:2014ooa}, usually also   referred to simply as non-metricity for short\footnote{Its definition is the following:
\bea
\!Q=\frac{1}{4}Q^{\alpha\beta\gamma}Q_{\alpha\beta\gamma}-\frac{1}{2}Q^{\gamma\alpha\beta}Q_{\alpha\gamma\beta}-\frac{1}{4}Q^\alpha Q_\alpha+\frac{1}{2}\tilde{Q}^\alpha Q_\alpha,
\label{Q}
\eea
where $Q_{\alpha}=Q_{\alpha\;\;\;\mu}^{\;\;\;\mu}$ and $\tilde{Q}_\alpha=Q^{\mu}_{\;\;\;\alpha\mu}$.}. 

To all intents and purposes such scalar encodes modifications on the length of a vector when it is transported parallely. Nevertheless, even though the degree of complexity of the theories arising from these assumptions is considerable, some recipes have been found leading to the derivation of new cosmological settings which offer exciting new prospects. But, obviously, in modified gravity settings it makes no sense to rely just upon an analytical sketch of the main evolutionary features related to their non-standard characteristics, it is also almost mandatory to perform observational  tests.

As it has been addressed {\it ad nauseam} in the literature, in order to  connect  theory and observations at the background level  the Friedmann and Raychaudhuri equations must be solved either explictly or implictly, with redshift typically being the parameter facilitating the connection. The procedure then progresses by confronting observationally either the Hubble factor itself or cosmological distances drawn from it. 

The typical extra levels of non-linearity in modified gravity as compared to GR lead sometimes to {\it a priori} not completely closed schemes in the context of the previous task, and then some ans\"atze may have to be  adopted that compromise generality somehow. Specifically, these are cases in which the piece $f(R,\dots)$ or simply $f(Q)$ in the modified gravity Lagrangian is postulated as a phenomenological function of redshift, thus leading to an expression of the Hubble factor in which the physical interpretation of the parameters is somewhat loose. 

Good news are  that some non-metricity settings are quite amenable to treatments which allow to present  the Hubble parameter  as an explicit expression of the energy density of the sources, and consequently of redshift (this is done following pertinent generalizations of otherwise standard procedures). The expressions thus obtained provide a clearer insight of the problem from the analytical perspective (all redshift regimes can be readily explored) and also bring added value for statistical (computational)  analyses of the evolution of the model under question, as it makes codes more efficient when combined with state-of-the-art cosmological probes. Interestingly, the GR limit  of these modified gravity frameworks will be easily recognisable, thus allowing for a neat statistical analysis.

Let us now present a summary of the organization of this manuscript. In section \ref{section:fqmodels} we will conveniently introduce theories of modified gravity with non-metricity. That very section includes an account of how  to construct genuine such settings based solely on non-metricity   which, however, do mimic GR flawlessly in all circumstances. This is opportune to   motivate  further steps.

In subsections \ref{subsection:dgp} and \ref{subsection:dgpish} we stretch the previous discussion to specific ``new" scenarios while keeping the GR equivalence as a particular case. Upon very general hypotheses we will be able to present two new versions of the Friedmann and Raychaudhuri equations which will eventually lead to exact expressions of the corresponding Hubble parameter. Some interesting and very general evolutionary features regarding accelerated expansion will be discussed, and  we will relate our models to some very well known cosmological models discussed in quite different theoretical frameworks. The specifications required to bring back our models to the $\Lambda$CDM standard will be pointed out.

Then we will move on to the statistical analysis section  \ref{tools}, based on the Markov Chain Monte Carlo (MCMC) procedure. The best fits, errors and statistical discriminators will be driven by supernovae type Ia luminosity data and modulated by direct Hubble (cosmic clocks), CMB shift and BAO data (head to section \ref{data_sketch} for further details). These data sets will allow to obtain tight enough constraints on the parameters, and results will be shown in section \ref{parameter_results}. Interestingly, the crop of the previous statistical analysis is of further use to produce constraints on derived parameters in the cosmographic territory, as we discuss in \ref{cosmodiagrams}. Lastly, we will dissertate about our main and secondary conclusions in section \ref{Conclusions}.

\section{$f(Q)$ models}
\label{section:fqmodels}
Throughout this work we will stick to a spatially flat Friedmann-Lemaître-Robertson-Walker (FLRW) spacetime, the golden standard that describes the Universe on large scales under the assumption of homogeneity and isotropy:
\bea
ds^2=-N^2(t)dt^2+a^2(t)\left[dx^2+dy^2+dz^2\right].
\eea
In order to explore the evolution of such spacetimes in the context of $f(Q)$ gravity (see \cite{BeltranJimenez:2018vdo,Jimenez:2019ovq, Lazkoz:2019sjl, Mandal:2020buf} for further details)  we must take into account that the  non-metricity scalar corresponding to the previous metric reads:
\bea
Q=6\frac{H^2}{N^2}.
\label{non-metricity scalar}
\eea
Following again \cite{BeltranJimenez:2018vdo,Jimenez:2019ovq}, we can take advantage of the fact  that $f(Q)$ theories allow to choose particular forms of the lapse function $N$ because $Q$ retains a residual time-reparametrization invariance in spite of some theoretical caveats, which allows to fix $N(t)=1$. 
In addition, applying variations with respect to the metric as usual, and following \cite{Xu:2019sbp, Jimenez:2019ovq}, 
one can reformulate the cosmological equations like:
\bea
6 f_Q H^2-\frac{1}{2}f&=&\rho, \label{Friedmann}\\
(12 H^2 f_{QQ}+f_Q)\dot H&=&-\frac{1}{2}(\rho+p) \label{Ray}.
\eea
Differentiation with respect to $Q$ is  denoted with its symbol as a subindex. In this fashion we will propose generic $b(Q)$ functions  that represent $\rho$ in Eq. \eqref{Friedmann}, so that the equation
\bea
Q f_Q-\frac{1}{2}f=b(Q), \label{defu}
\eea
can be integrated. As the latter is a first order non-homogeneous ordinary differential equation, it is not difficult to reduce its solution to a quadrature for any ansatz:
\bea f(Q)=\sqrt{Q}\int\frac{\displaystyle b(Q)}{Q^{3/2}}\eea
Although this formal result may in principle serve as a guidance to propose other ans\"atze, we just regard it as an anecdotal information, because only inspired enough function selections combined with Eq.  \eqref{Friedmann} render results which are  invertible and lead to expressions of $H^2$ as  explicit functions of $\rho$. Let us review the literature for a moment to address one such case. 

The starting point is to realize  that the GR background behaviour 
can be recovered under the prescription
\bea
b(Q)=\frac{Q}{16\pi G},\label{bqgr}
\eea
which implies
\bea
f(Q)=\frac{1}{8\pi G}\left(Q+M\sqrt{Q}\right)
\label{fqgr},
\eea
where $M$ is a constant which can be interpreted as a mass scale \cite{Jimenez:2019ovq}. 
It should not  be unanticipated that the $M=0$ case is analogous to GR, because it actually corresponds to the STEGR framework already discussed in quite a lot of detail  in \cite{Jimenez:2019ovq, Ayuso:2020dcu}. Note however, that the  $M\not =0$ choice leads to a  whole class of theories with the same background as GR with differences that can only be spotted at the level of perturbations.


Going back to the general procedure, it sets off by proposing a form of $b(Q)$ which, upon integration of Eq.  \eqref{defu}, produces an explicit expression of $f(Q)$ which we insert back into  Eq. \eqref{Friedmann}. Then, using Eq. \eqref{non-metricity scalar} we finally can relate $H^2$ to $\rho$ through an implicit equation. In general these equations will not be solvable for $H$, thus making it very complicated to progress significantly any further, as we discussed earlier. Fortunately, an array of cases can be found where this difficulty is overcome. 

It is the main result of this paper to present new two such scenarios and to examine them. Those two cases have been found by putting forward ans\"atze for $f(Q)$ which include Eq. \eqref{bqgr} as a particular case, and perhaps this may prove in the future a promising route. As we will  discuss the main connection between those two scenarios  is that non-metricity effects are screened in both of them. Besides, we will assume that on their respective GR limit they produce $\Lambda$CDM, the hardly beatable popular ``échantillon".


\subsection{DGP cosmologies from $f(Q)$ actions}
\label{subsection:dgp}

The choice $b(Q)\propto Q$ would be responsible for the appearance of the prescriptive $H^2$ term on the left hand side (lhs) of the Friedmann equation \cite{Jimenez:2019ovq,Ayuso:2020dcu}. For this reason it may be expected that  adding term proportional to $\sqrt{Q}$ will  make an $H$ term appear alonsgide $H^2$, thus leading to the sort of modified Friedmann equation characterizing several modified gravity scenarios. 

Let us show this is indeed the case by proposing
\bea
b(Q)=\frac{1}{16\pi G}\left( \alpha\sqrt{Q} +\beta Q\right)
\eea
in Eq. \eqref{defu}, where $\alpha$ and $\beta$ are  arbitrary constants. The latter can be integrated to give
\bea
f(Q)=\frac{1}{16\pi G}\left( \alpha\sqrt{Q}\log{Q}+2\beta Q\right),
\eea
where we have set to zero an integration constant which would give us a term proportional to $\sqrt{Q}$ again, which really does not induce any background dynamics. Using the form that the Friedman equation takes for this $f(Q)$ setting,
and translating the non-metricity scalar into the Hubble factor through the above mentioned and known relation $Q=6 H^2$, we arrive at
\bea
H \left(\sqrt{6}\alpha + 6 \beta H\right)=16\pi G\rho.
\eea
Alternatively, if we opt for a more enlightening rendering of the same relation, we can write
\bea
H^2+\frac{\alpha}{\sqrt{6}\beta}H=\frac{8\pi G}{3\beta}\rho.\label{H1}
\eea
An exact GR framework is recovered for the simultaneous choice $\alpha=0$, $\beta=1$ (see Eq. \eqref{H1}). We solve either of the former two expressions for $H$ and select the  branch that corresponds to an expanding universe ($H>0$) to conclude
\bea
H=\frac{\sqrt{6}}{12\beta}\left(\sqrt{\alpha^2+ 64\pi G \beta \rho }-\, \alpha\right).
\eea

Incidentally, the background evolution of this model for $\beta=1$   is just identical to that of the well-known DGP cosmologies \cite{Deffayet:2000uy} studied in a huge collection of references, such as for instance in \cite{Lazkoz:2006gp}. The $\alpha<0$ case is self accelerated (acceleration will occur even if its only matter content would not produce acceleration on it own, for instance cosmic dust). The opposite applies to the $\alpha>0$ case. We will try to explore both models simultaneously as long as it is possible, while  fixing $\beta=1$ (this must be stressed). 

Let us  now make the typical assumption in other references studying $f(Q)$ cosmologies that the matter-energy content is a sum of barotropic fluids $\rho=\sum\rho_i$, and that each one of them satisfies separately the standard conservation equation:
$\dot \rho_i+3H(1+w_i)\rho_i=0,$
where on the one hand the additional assumption of simple equations of state mediated  by constant  $w_i$ parameters has been made as well and the dot means time derivative.

From  the discussion in \cite{Lazkoz:2006gp} some easy conclusions about the behaviour of the models can be drawn.  To this end we consider a combination of three  fluids:  cosmic dust, a cosmological constant and radiation. In that case thus $\rho=\rho_m+\rho_r+\rho_{\Lambda}$, and we will also assume $\rho_m,\; \rho_r,\; \rho_{\Lambda} \ge 0$, which will be relevant for the reminder. Under this assumption, and upon differentiation of Eq. (\ref{H1}), we obtain
\bea 
\dot H=-\frac{4\pi G}{3}(3\rho_m+4\rho_r) \left(1-\frac{\alpha }{\sqrt{\alpha ^2+64\pi G \rho }}\right)
\eea
It is clear that $\dot H
<0$ necessarily for whatever sign of $\alpha$.

As the universe expands  the linear term in $H$ appearing in Eq. (\ref{H1}) ceases to be negligible as compared to the quadratic one, and the $f(Q)$ effects cannot be waived any longer. Actually, a screening of the cosmological constant arises and  an effective dark energy turns out to offer a good description:
\bea
\rho_{\rm eff}=\rho_{\Lambda}-\frac{  \sqrt{6} \alpha  H}{16 \pi  G} \label{rhoeffdef}.
\eea
Here $\rho_{\rm eff}$ offers a convenient way to encode the modification in a general relativistic fashion which, together with an effective equation of state parameter $w_{\rm eff}$, allows us to write the following couple of equations:
\bea
H^2=\frac{8\pi G}{3}\left(\rho_m+\rho_r+\rho_{\rm eff}\right),\label{heff}\\
\dot \rho_{\rm eff}+3H(1+w_{\rm eff})\rho_{\rm eff}=0.\label{conseff}
\eea

Combining Eq. (\ref{conseff}) and Eq. (\ref{rhoeffdef}) we arrive at
\bea
(1+w_{\rm eff})=\frac{  \sqrt{6} \alpha  \dot H}{48 \pi  G \rho_{\rm eff}H},
\eea
which can be evaluated using the reformulation
\bea
\rho_{\rm eff}=\rho_{\Lambda}+\frac{\alpha}{32 \pi  G}\left(\alpha-  \sqrt{\alpha ^2+64 \pi  G \rho}\right)\;.
\eea
It can be seen that for $\alpha<0$ a phantom behaviour is excluded (as a consequence of $\rho_{\rm eff}$ being positive, which gets translated into $w_{\rm eff}>-1$), whereas for  $\alpha>0$ there is no such guarantee.

The following necessary step consists in recasting our expressions as functions of redshift, thus paving the way for an observational analysis. Given the hypotheses we made earlier we render the total energy density $\rho$ in the following form:
\bea
\rho=\frac{3H_0^2}{8\pi G}\left[\Omega_m(1+z)^3+\Omega_{\Lambda}+\Omega_r(1+z)^4\right],
\label{rhoz}
\eea where we have introduced the following implicit definitions:
\bea
&&\rho_m=\frac{3H_0^2\Omega_m}{8\pi G}(1+z)^3, \label{omegam}\\
&&\rho_r=\frac{3H_0^2\Omega_r}{8\pi G}(1+z)^4, \label{omegar}\\
&&\rho_{\Lambda}=\frac{3H_0^2\Omega_{\Lambda}}{8\pi G}, \label{omegalambda}
\eea
as well as \footnote{Alternative definitions of this parameter so that dimensionally/aesthetically would stand on the same grounds as, say, $\Omega_{\Lambda}$ will force the need to consider sign duplicities
and square roots which will induce less transparency in the geometry of the space of parameters and will complicate unnecessarily the codes to perform observational tests.}
\bea
\Omega_Q=\frac{\alpha}{2\sqrt{6} H_0}, \label{omegaQ}
\eea
so that we  arrive at the expression
\bea
E(z)=\sqrt{ \Omega_\Lambda+\frac{8\pi G}{3H_0^2}(\rho_m+\rho_r) +\Omega_Q^2}-\Omega_Q\qquad \label{Efunbrane}
\eea
upon the prescription $H(z)=H_0E(z)$. It is obvious then that  the following applies  for $H$ as given by Eq. (\ref{Efunbrane}):
\bea
\lim_{\Omega_Q\to 0}H^2=H_0^2 \left[\Omega_{\Lambda} + \Omega_m (1 + z)^3+\Omega_r(1+z)^4\right]. \label{limitH}
\;\quad\eea

Furthermore, from Eq. (\ref{Efunbrane}) it can be seen that the non-metricity features encoded in the parameter $\Omega_Q$ are somewhat screened: explicitly, the choice $\Omega_m=\Omega_{\Lambda}=\Omega_r=0$ gives a null $H$ function for any value of $\Omega_Q$ whatsoever.

On the other hand, the  customary
normalization condition $E(z=0)=1$ enforces  an  extra condition:
\bea
\Omega_Q=\frac{1}{2}(\Omega_m+\Omega_\Lambda+\Omega_r-1).
\eea


Note that (from Eq. (\ref{Efunbrane}) once again)  a positive $\Omega_Q$ value will slow down the expansion as compared to the  $\Lambda$CDM case, whereas  a negative one will exert the contrary effect. An alternative way to see this is that for $\alpha<0$ the effective dark energy term is bigger than a bare  cosmological constant, as Eq. (\ref{rhoeffdef}) suggests.
In any case, an observational inference of a large absolute value of $\Omega_Q$  would be extremely unexpected if we take into account  the mounting evidence of a universe extremely agreeable with the $\Lambda$CDM behaviour.

\subsection{DGPish  cosmologies from $f(Q)$ actions}
\label{subsection:dgpish}
We may explore other  routes compatible with a $b(Q)\propto Q$ behaviour in an appropriate limit and thus leading to the standard $H^2$ term on the left hand side of the Friedmann equations, while reproducing a different behaviour in other regimes.
We propose now a new case inspired by our first case, which
explictly stems from the assumption 
\bea
b(Q)=\frac{1}{16\pi G}\sqrt{\gamma Q + \beta^2 Q^2}
\eea
where $\gamma$ and $\beta$ are, in principle, two arbitrary constants, even though from early lessons we may anticipate that we will have to fix $\beta=1$ along the road. Upon integration of Eq. (\ref{defu}) we arrive at:
\bea
f(Q)=&&\frac{Q \sqrt{u(Q)} \left(\sqrt{u(Q)}-\sqrt{\gamma } \arctanh \left(\frac{\sqrt{u(Q)}}{\sqrt{\gamma }}\right)\right)}{8 \pi  G
   \sqrt{Q u(Q)}}\qquad
\eea
where $u(Q)=\gamma +\beta ^2 Q$.
Keeping our discussion general for the time being, and following the same recipe as before, we obtain
\bea
H \sqrt{\gamma + 6 \beta^2 H^2}=\frac{16\pi G}{\sqrt{6}}\rho.
\eea
This  can be solved for $H$ to write
\bea
H^2=\frac{\sqrt{\gamma^2+(32\pi G\beta\rho)^2}-\gamma}{12 \beta^2}.
\label{hsquarenewmodel}
\eea
Besides, if  the expanding branch is chosen, we can go further and write
\bea
H=\frac{\sqrt{\sqrt{\gamma^2+(32\pi G\beta\rho)^2}-\gamma}}{2\sqrt{3}\,\beta}.\label{hnewmodel}
\eea

We can follow the same sequence of steps as for the first case so as to throw some light on the evolution of this model, where once again we assume $\rho=\rho_m+\rho_r+\rho_{\Lambda}$ and $\beta=1$. Under this assumption and upon differentiation of Eq. (\ref{hsquarenewmodel}) we obtain
\bea 
\dot H=-\frac{128 \pi ^2 G^2(3 \rho_m+4 \rho_r)\rho }{3 \sqrt{\gamma ^2+1024 \pi ^2 G^2 \rho^2}},
\eea
which is definite negative under our hypotheses on the positivity of  the various  densities, regardless of the  sign of $\gamma.$ In this case, finding the explicit expression of $\rho_{\rm eff}$ is not so straightforward, but we can see that it reads
\bea
&&\rho_{\rm eff}=\rho_{\Lambda}+\frac{3 H}{8 \pi  G}\left(H-\frac{ \sqrt{\gamma +6H^2}}{\sqrt{6}}\right).
\label{rhoeffdgpish}
\eea

We then process our definitions and equations and eventually get
\bea
(1+w_{\rm eff})=
\frac{\dot H\sqrt{6}\left[\sqrt{\gamma^2+(32\pi G \rho)^2}-32\pi G \rho\right]}{48 \pi  G \rho_{\rm eff}H\sqrt{\gamma +6 H^2}}\ ,
\eea
which can be evaluated using the reformulation
\bea
\rho_{\rm eff}=\rho_{\Lambda}-\rho+\frac{\sqrt{\gamma ^2+(32\pi G)^2 \rho^2}-\gamma}{32 \pi  G}.
\eea

It is clear from Eq. \eqref{rhoeffdgpish} that for positive $\rho_{\Lambda}$ and $H$ we can guarantee $\rho_{\rm eff}>0$ if $\gamma<0$ (for the case with $\gamma>0$ is not possible to conclude the sign of $\rho_{\rm eff}$), and whatever the sign of $\gamma$ it follows that  $\dot H
<0$. Consequently, we can conclude that $1+w_{\rm eff}<0$ for $\gamma<0$ necessarily.


Before we proceed, let us remind that not all choices of $b(Q)$ that can be thought of will lead to an invertible relation between $H$ and $\rho$, and therefore their potential use will get compromised in a way. Having made this comment, let us go on building from either Eq. (\ref{hsquarenewmodel}) or Eq.
(\ref{hnewmodel}) by choosing again Eq.  
(\ref{rhoz})
as our expression for $\rho$ as a function of redshift.

Following the same steps as before 
and by  sticking with the general case
we can write
\bea
E(z)= \sqrt{\sqrt{\left(\Omega_\Lambda+\frac{8\pi G}{3H_0^2}(\rho_m+\rho_r)\right)^2+\Omega_Q^2}-\Omega_Q},~~~~
\eea
where now 
\bea
\Omega_Q=\frac{\gamma}{12 H_0^2}\ ,
\eea
and from the normalization, we obtain
\bea
\Omega_Q=\frac{1}{2}\left[(\Omega_\Lambda+\Omega_m+\Omega_r)^2-1\right].
\eea

Again, for our new expression describing $H^2$  we see that the limit of Eq.
(\ref{limitH}) 
happens here too recovering GR for $\gamma=0$. 



\section{Observational data and statistical analysis}

\subsection{Basic statistical tools}\label{tools}
In order to fit the free parameters of our modified gravity models we will use several samples of observational data. In addition, and within the same statistical framework,  we will fit the parameters of $\Lambda$CDM for comparison. This will allow to drawing conclusions about the statistically admissibility of the new models under study. Specifically, as customarily established, we will find
 the values of the free parameters which maximize the posterior (probability density)  produced by combining a prior (probability density) with hypothetical Gaussian likelihoods associated with  $\chi^2$ functions built for each dataset and then a combined one obtained through the product of likelihoods or the sum of $\chi^2$ (of course, priors can be combined as well). But, strictly speaking,  our ``best fits" here will be drawn from the median of the posterior so that we account for deviations from perfect Gaussianity, which will also reflect in asymmetrical errors. So even though we will loosely talk about ``best-fits", the remark we just made will have to be taken into account at all times.

For homogeneity,  we choose priors that at are applicable to both $\Lambda$CDM and to the two modified gravity models \footnote{Let us remark  that spatially flatness   implies $\Omega_m+\Omega_r+\Omega_\Lambda=1$ in $\Lambda$CDM.}:
\begin{itemize}
\item $0\leq\Omega_m+\Omega_\Lambda+\Omega_r$,
\item $0<(\Omega_m,\Omega_\Lambda,h)<1$,
\item $\Omega_m>\Omega_b>\Omega_r$.
\end{itemize}

In addition, one could explore the reliability of the model by means of the evidence $\mathcal{E}$, which is the weighted average of the likelihood with the prior(s) acting as weights. This quantity is often regarded as the best founded statistical tool in order to compare models in cosmology \cite{Nesseris:2012cq}, provided that wide-enough priors are chosen. Thus, to compare two different models one should introduce the Bayes factor, which is the ratio of evidences of the models:
\bea
\mathcal{B}^{i}_{j}=\frac{\mathcal{E}_i}{\mathcal{E}_j}.
\eea
and consequently if ${\mathcal{E}_i}>{\mathcal{E}_j}$  then $\mathcal{B}^{i}_{j}>1$  for the measured data, and from this it follows that model $\mathcal{M}_i$ will be preferred over  model $\mathcal{M}_j$.
However, the next question is how much statistically preferred one model over the other. To help us discern it, we have Jeffreys' Scale \cite{Jeffreys61} whose criterion is the following: if $\ln \mathcal{B}^{i}_{j}<1$, the evidence in favor of the model $\mathcal{M}_i$ is not significant; if $1<\ln \mathcal{B}^{i}_{j}<2.5$, the evidence is substantial; if $2.5<\ln \mathcal{B}^{i}_{j}<5$, it is strong; and if $\ln \mathcal{B}^{i}_{j}>5$, it is decisive. Along this work, we will fix $\mathcal{E}_j$ as the evidence of $\Lambda$CDM to be used as a standard comparative value and $\mathcal{E}_i$ will be the evidence of the studied models.

Some specifics of the four data sets we will put into consideration are given in the next subsection.
The reason why we will explore individual $\chi^2$ values and their best fits is the possibility to estimate the contribution of each data set to the final result and conclusions. Tables to summarize our findings will be presented. In addition, complementary and very relevant information will be provided through confidence contours addressing again the separate and joint data set analyses.

\subsection{Datasets} \label{data_sketch}
\subsubsection{Pantheon Supernovae data}
The Pantheon sample \cite{Scolnic:2017caz}, made of  1048 SNeIa data with its distance modulus in the redshift range $0.01<z<2.26$. Their characteristics allow them to be used as standard candles and from their distance moduli we can extract  luminosity distances. In this way we can build a  $\chi^2$ function to set theoretical predictions to the test. 

\subsubsection{Hubble data}

This is a sample of $31$ values of the Hubble function, $H(z)$, in the redshift range $0.07<z<1.965$ \cite{Moresco:2016mzx, Moresco:2015cya}. They are obtained from early  time passively evolving galaxies for which an estimate of their age can be obtained due to to some peculiar spectral features. This makes them represent reliable cosmic chronometers which in turn allow to construct a  $\chi^2$ function to constraint free parameters in our theoretically proposed expression for $H(z)$.

\subsubsection{CMB}

In this case we will use information extracted from cosmic microwave background (CMB) data which can be translated into the  so-called three shift parameters \cite{Wang:2015tua}, thus allowing to explore the evolution of the cosmological background.
This set of three quantities takes advantage of the possibility to estimate precisely how sensitive the CMB is to the distance
to the decoupling epoch as encoded in the locations of peaks and
troughs of the acoustic oscillations. Specifically there are two measurable quantities. The first one is the angular diameter distance to the
decoupling epoch divided by the sound horizon size at
the decoupling epoch, which forms the so called acoustic scale. The second is the quotient between the angu-
lar diameter distance to the decoupling epoch divided
and the Hubble horizon size at the decoupling epoch. As customary in the literature \cite{WMAP:2008lyn,Zhai:2019nad},  these two distance priors are  combined with the normalized density fraction of baryons to  construct a data vector and then  the corresponding $\chi^2$ estimator. 

\subsubsection{BAO}
Baryon Acoustic Oscillations (BAO) are a pattern of  oscillations drawn from the physics of fluctuations in the density of visible baryonic matter produced by acoustic density waves in the primordial plasma. This fluctuations lead to measurable features in the distribution of galaxies that can be used as standard rules, calculating the maximum distance that acoustic waves  can reach before the plasma cooled at the recombination moment when the wave is frozen.

This set of data is composed in five subsets from different missions: the WiggleZ Dark Energy survey \cite{Blake:2011en}, the SDSS-III Baryon Oscillation Spectroscopy Survey (BOSS) DR12  described in \cite{Alam:2016hwk}, the  extended Baryon Oscillation Spectroscopy Survey (eBOSS), the Quasar-Lyman $\alpha$ Forest from SDSS-III BOSS DR11 \cite{Agathe:2019vsu},and the Voids-galaxy cross-correlation data from \cite{Nadathur:2019mct}.


{\renewcommand{\tabcolsep}{1.5mm}
{\renewcommand{\arraystretch}{1.5}
\begin{table*}[h!]
\caption{MCMC best fits and errors along with other statistical estimators. Quantities in italic correspond to secondary  parameters.}
\begin{center}
\begin{tabular}{ccccccc}
\hline
\hline
& & Pantheon & Hubble & CMB & BAO  & Total  \\
\hline 
\multirow{3}{*}{$\Omega_m$} & $\Lambda$CDM &  $0.297_{-0.023}^{+0.023}$   & $0.328_{-0.056}^{+0.065}$ & $0.316_{-0.007}^{+0.007}$& $0.320_{-0.015}^{+0.016}$& $0.323_{-0.005}^{+0.005}$\\
\multicolumn{1}{l}{}       &    $DGP$   &   $0.277_{-0.051}^{+0.048}$   &  $0.316_{-0.065}^{+0.076}$ &   $0.358_{-0.053}^{+0.057}$   &  $0.301_{-0.018}^{+0.019}$ &  $0.326_{-0.006}^{+0.006}$   \\
\multicolumn{1}{l}{}       &    $DGPish$   &  $0.307_{-0.026}^{+0.027}$  &  $ 0.326_{-0.055}^{+0.066}$ & $0.310_{-0.014}^{+0.009}$ &   $0.320_{-0.015}^{+0.016}$ & $0.322_{-0.005}^{+0.005}$  \\
\hline
\multicolumn{1}{c}{\multirow{3}{*}{$\Omega_b$}} & $\Lambda$CDM &  $-$ &  $-$  &$0.0490_{-0.0006}^{+0.0006}$ & $0.057_{-0.024}^{+0.010}$ &  $0.0496_{-0.0004}^{+0.0004}$\\
\multicolumn{1}{l}{}  & $DGP$ &  $-$  &  $-$ &  $0.0557_{-0.0083}^{+0.0089}$ &  $0.066_{-0.027}^{+0.013}$&   $0.051_{-0.001}^{+0.001}$       \\
\multicolumn{1}{l}{}       &    $DGPish$   &    $-$   &   $-$ &
$0.0483_{-0.0021}^{+0.0009}$  & $0.056_{-0.025}^{+0.010}$ &  $0.0494_{-0.0005}^{+0.0005}$ \\
\hline
\multicolumn{1}{c}{\multirow{3}{*}{$h$}} & $\Lambda$CDM &  $-$  & $0.677_{-0.031}^{+0.031}$ & $0.675_{-0.005}^{+0.005}$& $0.70_{-0.15}^{+0.18}$ &  $0.670_{-0.003}^{+0.003}$ \\  \multicolumn{1}{l}{}       &    $DGP$     & $-$  &   $0.673_{-0.031}^{+0.030}$    &   $0.634_{-0.045}^{+0.053}$  & $0.69_{-0.16}^{+0.17}$ & $0.665_{-0.006}^{+0.006}$     \\
\multicolumn{1}{l}{}       &    $DGPish$   & $-$   &  $0.682_{-0.032}^{+0.032}$ &  $0.681_{-0.008}^{+0.015}$  & $0.70_{-0.16}^{+0.18}$ &  $0.671_{-0.004}^{+0.004}$ \\ 
\hline
\multicolumn{1}{c}{\multirow{3}{*}{$\Omega_\Lambda$}} & $\Lambda$CDM &
${0.702_{-0.023}^{+0.022}}$ &  ${0.672_{-0.065}^{+0.056}}$ & ${0.684_{-0.007}^{+0.007}}$ & ${0.680_{-0.016}^{+0.015}}$ &  ${0.677_{-0.005}^{+0.005}}$ \\
\multicolumn{1}{l}{}    &    DGP       & $0.58_{-0.32}^{+0.27}$  & $0.53_{-0.31}^{+0.30}$ & $0.43_{-0.28}^{+0.32}$ & $0.38_{-0.15}^{+0.18}$ &   $0.629_{-0.047}^{+0.047}$       \\
\multicolumn{1}{l}{}       &    DGPish   &   $0.53_{-0.31}^{+0.30}$  &  $0.50_{-0.31}^{+0.32}$ &  $0.51_{-0.31}^{+0.30}$  & $0.64_{-0.19}^{+0.19}$ &  $0.64_{-0.19}^{+0.19}$  \\
\hline
\multicolumn{1}{c}{\multirow{3}{*}{$\Omega_Q$}} & $\Lambda$CDM &  $-$  & $-$  & $-$ & $-$ & $-$ \\
\multicolumn{1}{l}{}  & DGP  & $\mathit{-0.07_{-0.19}^{+0.16}}$  & $\mathit{-0.08_{-0.17}^{+0.17}}$ & $\mathit{-0.10_{-0.11}^{+0.13}}$  & $\mathit{-0.161_{-0.082}^{+0.093}}$ &      $\mathit{-0.022_{-0.022}^{+0.022}}$ \\
\multicolumn{1}{l}{}       &    DGPish  &   $\mathit{-0.16_{-0.19}^{+0.29}}$   &  $\mathit{-0.16_{-0.22}^{+0.32}}$ &  $\mathit{-0.16_{-0.22}^{+0.29}}$  & $\mathit{-0.04_{-0.17}^{+0.20}}$ &  $\mathit{-0.04_{-0.16}^{+0.20}}$ \\
\hline 
\hline
\multicolumn{1}{c}{\multirow{3}{*}{$\chi^2_{\rm bf}$}} & $\Lambda$CDM & 1035.77 & 14.49 & 0.0013 &  16.55 &  1072.19 \\
\multicolumn{1}{l}{}                          & $DGP$   & 1035.75 & 14.50 & 0.0022 & 13.58&  1071.20 \\ 
\multicolumn{1}{l}{}                          &    $DGPish$   & 1035.77 & 14.33 & 0.016 & 16.55 & 1072.22 \\
\hline
\multicolumn{1}{c}{\multirow{3}{*}{$\mathcal{B}^{i}_{j}$}} & $\Lambda$CDM & $-$ & $-$ & $-$ & $-$ & 1 \\
\multicolumn{1}{l}{}                                      & DGP & $-$ & $-$ & $-$ & $-$ & 1.16\\
\multicolumn{1}{l}{}                                      & DGPish & $-$ & $-$ & $-$ & $-$ & 0.83 \\
\hline
\multicolumn{1}{c}{\multirow{3}{*}{$\ln \mathcal{B}^{i}_{j}$}} & $\Lambda$CDM & $-$ & $-$ & $-$ & $-$ & 0 \\
\multicolumn{1}{l}{}                                      & $DGP$ & $-$ & $-$ & $-$ & $-$& $0.15$\\
\multicolumn{1}{l}{}                                      & $DGPish$ & $-$ & $-$ & $-$ & $-$ & $-0.18$ \\
\hline
\end{tabular}
\end{center}
\label{tableresults}
\end{table*}}}

\begin{figure*}[htbp]
    \includegraphics[width=1.0\linewidth]{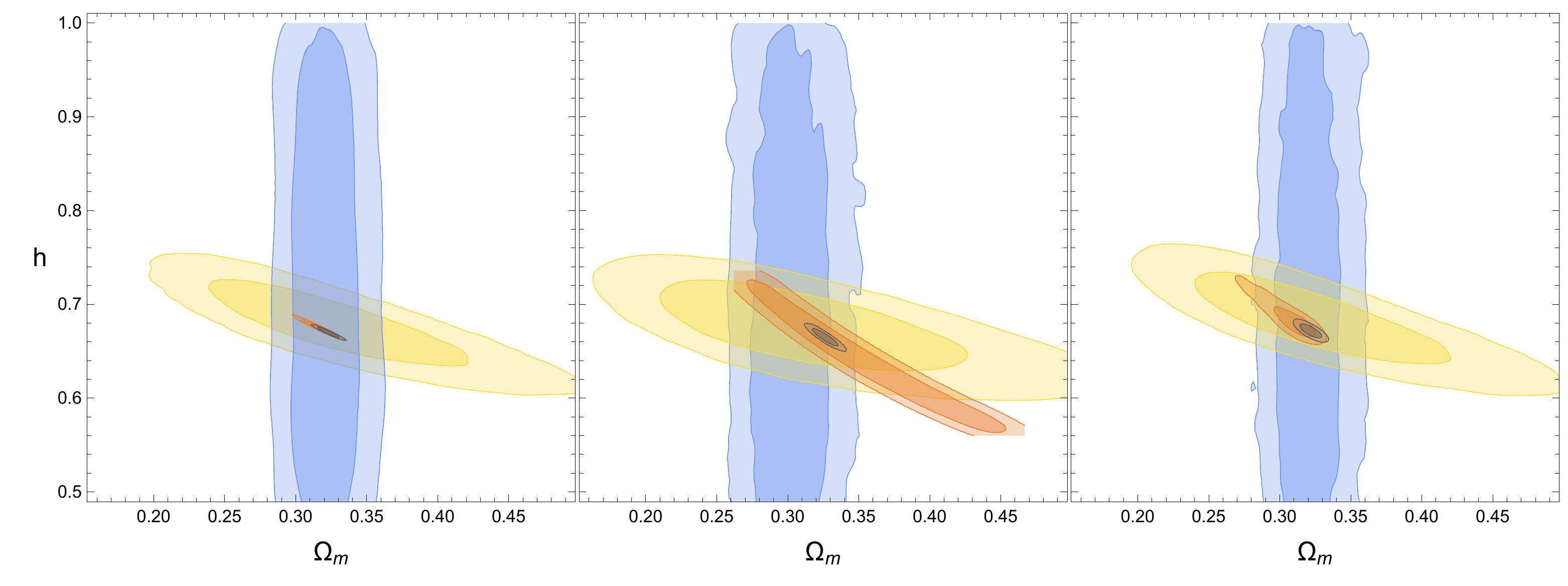}\\
    \includegraphics[width=1.0\linewidth]{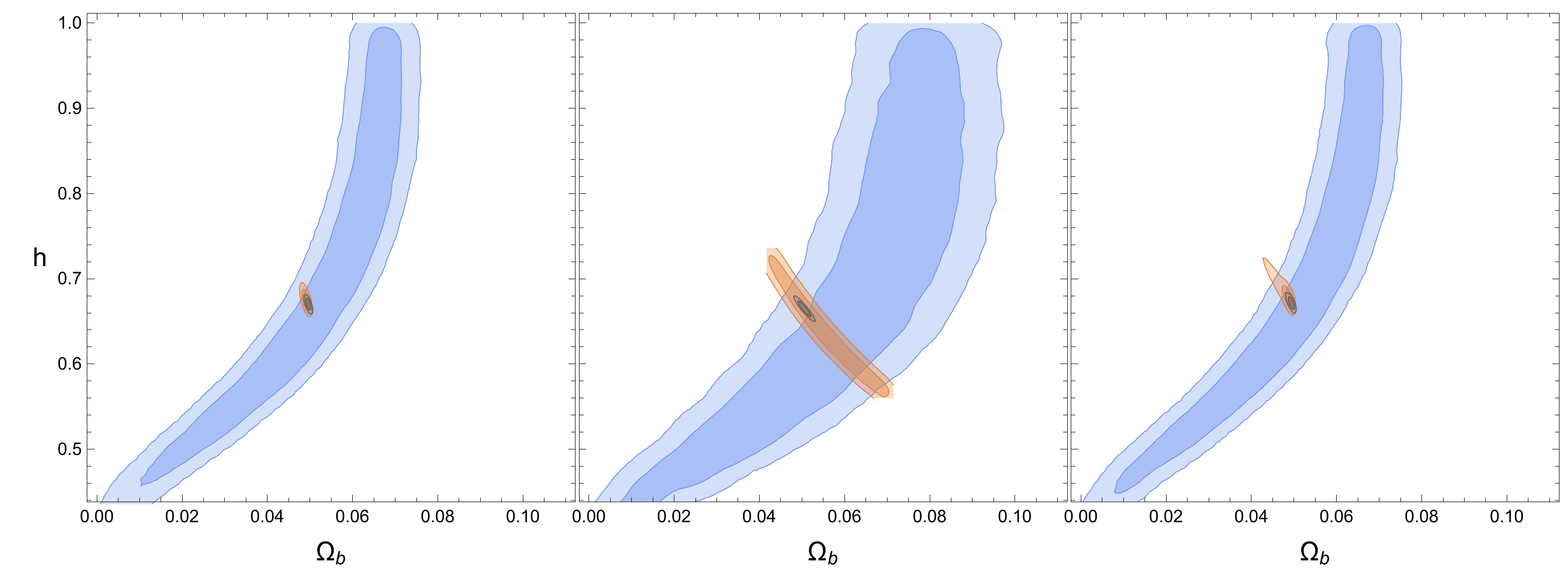}\\
    \includegraphics[width=1.0\linewidth]{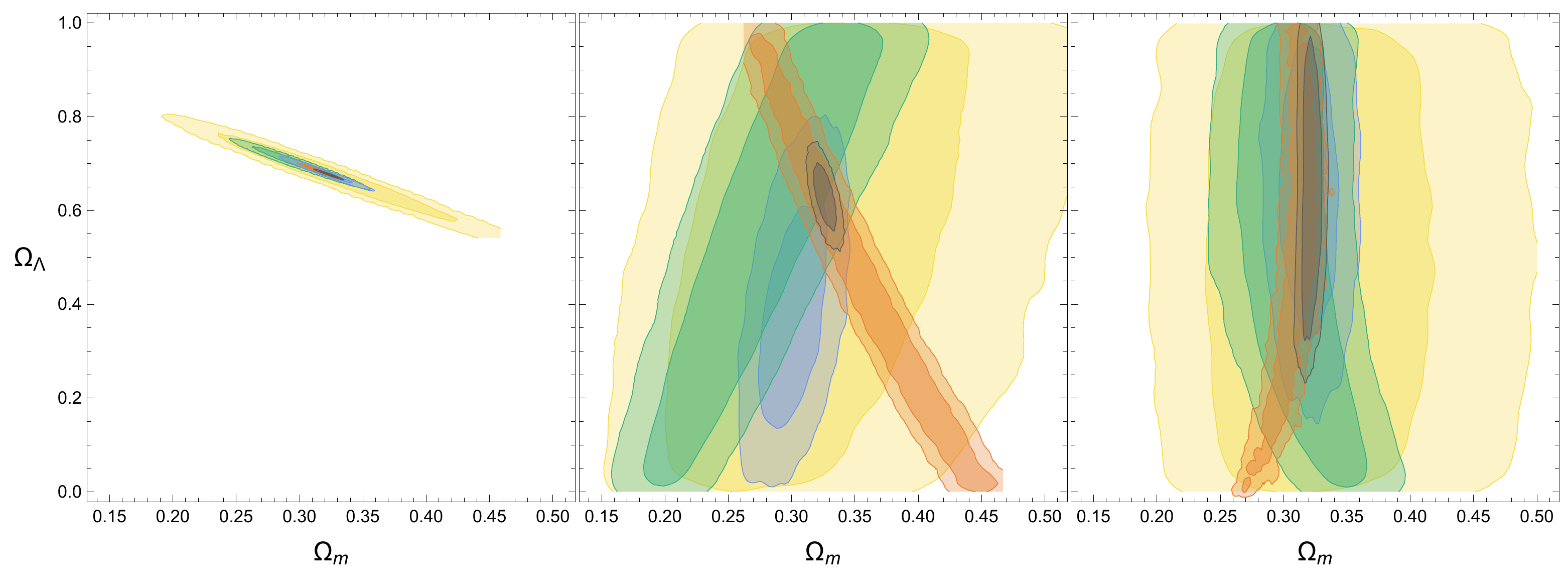}
    \caption{Confidence contours for the $\Lambda$CDM model (left column) $DGP$ model (center column) and $DGPish$ model (right column) with the following color scheme: green  - SNeIa, yellow - Hubble data, orange - \textit{Planck 2018} CMB, blue - BAO data, black - all sets of data. As SNeIa are not (of course) able  to  fix the  value $h$ their contours are missing from those plots where constraints on $h$ are represented; for the same rationality, both SNeIa and Hubble data contours are absent from plots showing constraints on $\Omega_b$.}
     \label{3Contours}
\end{figure*}

\begin{figure*}[htb]
    \includegraphics[width=0.9\linewidth]{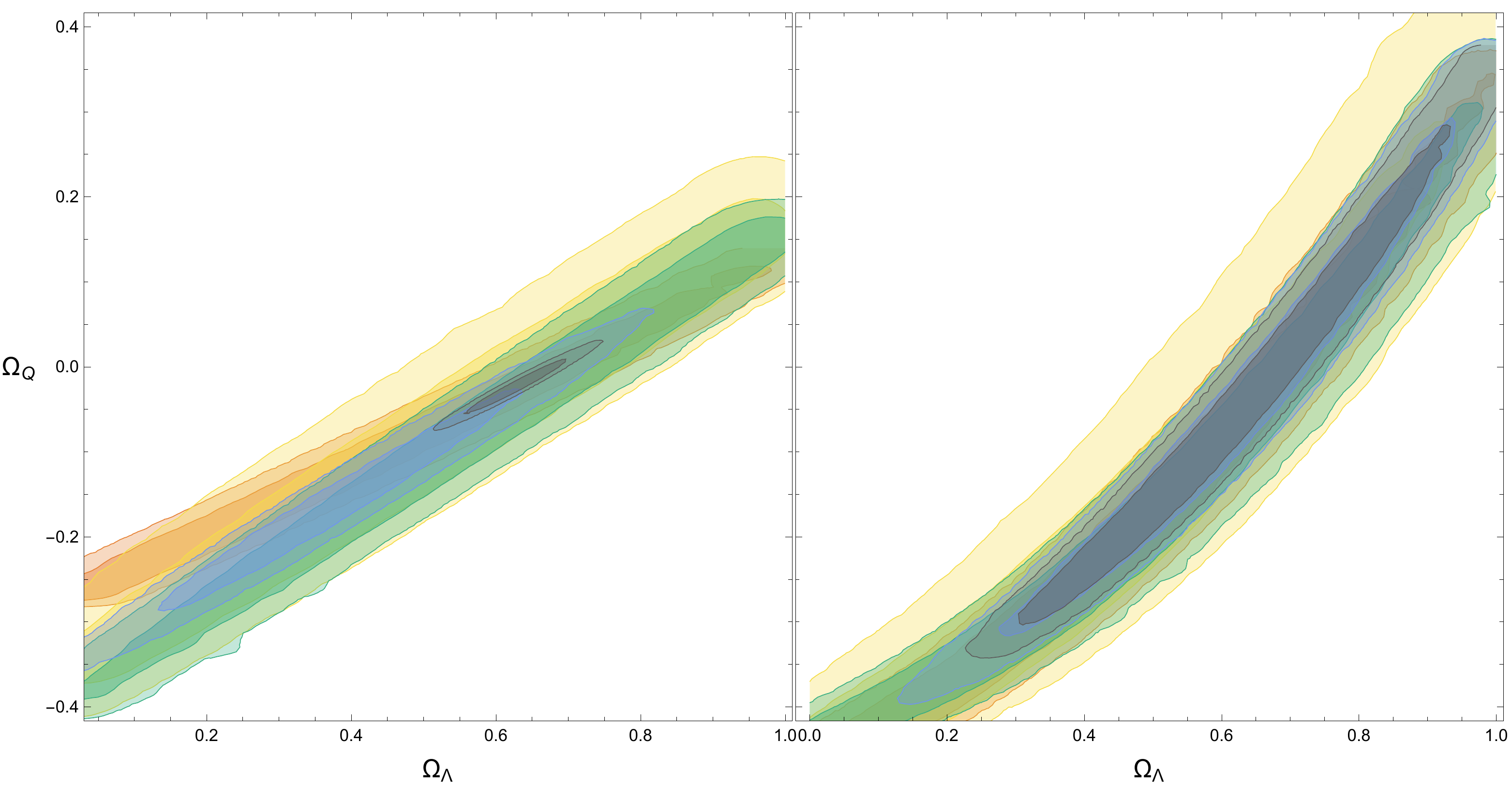}~~~\\
   \includegraphics[width=0.9\linewidth]{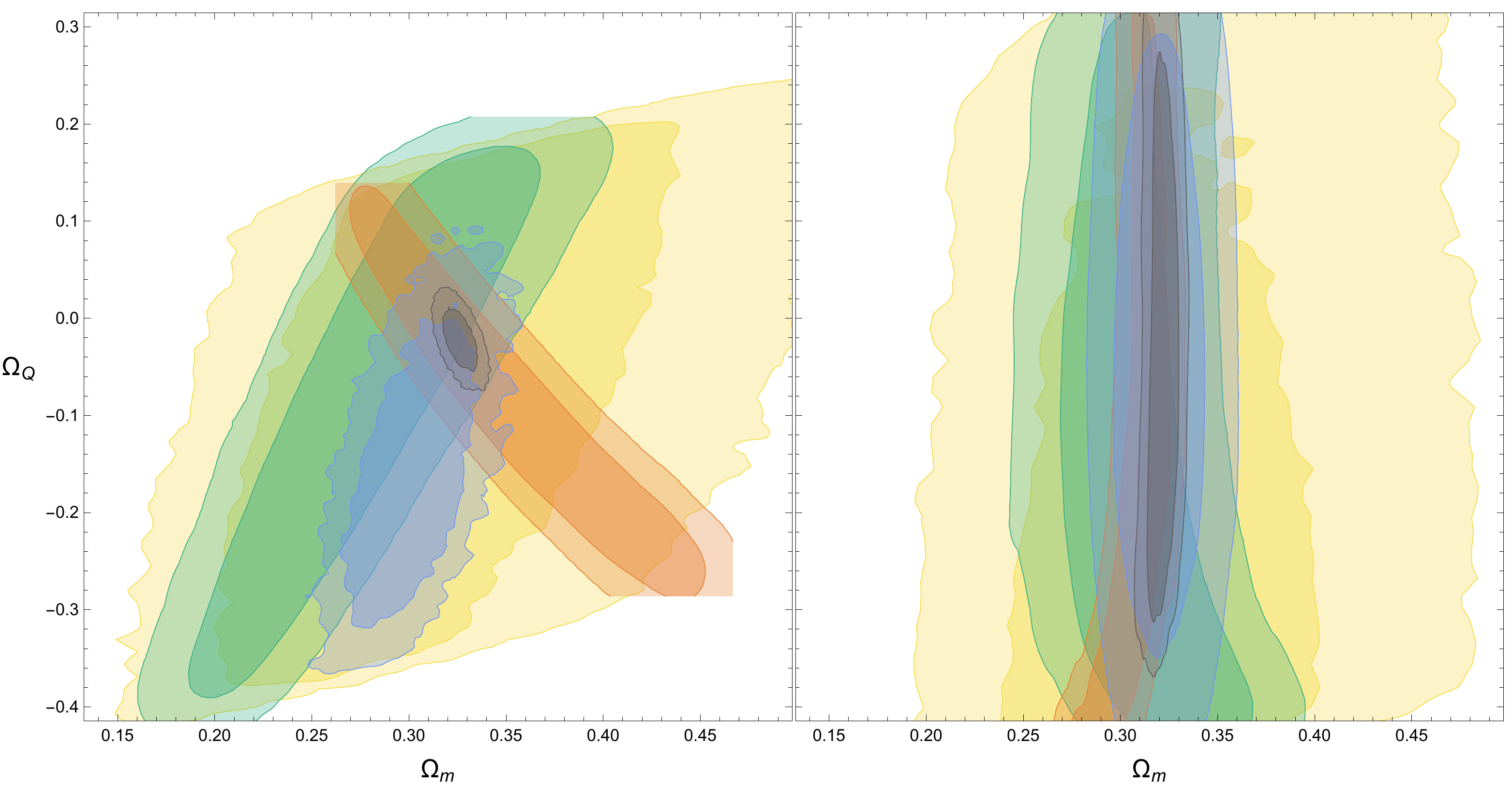}~~~
\caption{Contour plots of the constraints on the parameter $\Omega_Q$ for the $DGP$ model (left column) and $DGPish$ model (right column) with the following color scheme: green  - SNeIa, yellow - Hubble data, orange - \textit{Planck 2018} CMB, blue - BAO data, black - all sets of data. }
    \label{figOQ}
\end{figure*}

\subsection{Constraints on parameters} \label{parameter_results}

Having presented the statistical course of action and having chosen the astrophysical probes, we can set forth the results of our analysis. Throughout the discussion we will  often  compare our findings to those in the $\Lambda$CDM case. This  will be made possible by performing all the tests also on this scenario. The corresponding best fits along with those of our two $f(Q)$ models are shown on Table \ref{tableresults}. Note that the $\Omega_Q$ parameter is a signature of the $f(Q)$ scenarios alone.

For additional insight, we use the MCMC procedure (which yields our best fits) to obtain a convenient byproduct: confidence contours. They give us visual indications as to data sets complementarity, tightness of constraints, and  parameter correlations. In the corresponding plots   (Figs. \ref{3Contours}-\ref{figOQ}) we choose two different hues of several single colours to represent the $1\sigma$ (dark) and $2\sigma$ (light) regions, and then we associate each single colour to an  individual data set or data set combination (see figure captions).

For the most standard parameters, that is,  $\Omega_m$, $\Omega_b$ and $h$, we conclude (mainly from  Table \ref{tableresults} and Fig. \ref{3Contours}) that best fits,  uncertainty and data set complementarity are very much the same in the three scenarios ($\Lambda$CDM, DGP and DGPish)
. The most noticeable discrepancies arise when CMB data are considered alone, but for the combination of data sets differences are minimal. Stronger  disagreements arise (mainly in the size of errors) in $\Omega_{\Lambda}$ due to its blurring with the new parameter $\Omega_Q$.   In any case,  it would be worth exploring the effects of  generalizations of our models with additional parameters, in particular if the generalizations gives non-phantom models; but  we this leave for future prospects.

Now, even though $\Omega_{\Lambda}$ belongs in the classical category of parameters, we must consider it separately due to its non-trivial mixing with the new parameter.  From the third row of Fig. \ref{3Contours} we notice that $\Omega_{\Lambda}$ displays more uncertainty in the modified gravity models than in $\Lambda$CDM, which again, we put down to the fact that $\Omega_Q$ acts as $\Omega_{\Lambda}$ somehow, which is otherwise obvious from the Hubble function. 
In the DGPish case the uncertainty is way bigger than in the other two cases, there is practical no correlation whatsoever among $\Omega_{\Lambda}$ and $\Omega_m$ as shown in the contours, and the huge size of the errors makes us declare $\Omega_{\Lambda}$ is basically unconstrained in this case 
.

Along the same vein, the upper row on Fig. \ref{figOQ} tells us there is indeed a high correlation between $\Omega_Q$ and $\Omega_{\Lambda}$. Note, however that the combination of all data sets gives a far smaller contour in the DGP case, and therefore, as we have already stated $\Omega_Q$ is better fitted. 


The conclusions on constraints on $\Omega_Q$ and its role as related to $\Omega_{\Lambda}$ are strengthened by the contour which   confronts $\Omega_Q$ with $\Omega_m$ (see Fig. \ref{figOQ} again). In the DGPish case we get a very large almost vertical contour, that is, almost the narrow fit in $\Omega_m$ is compatible with a very wide range of $\Omega_Q$ values, that is  the modified gravity parameter $\Omega_Q$ is almost blind to the matter density. In contrast in the DGP case constraints are much more significant.

\section{Cosmodiagrams}\label{cosmodiagrams}

Determining the values of parameters in the so-to-say ``cosmographic spectrum" is quite a relevant task given their implications for the evolution and final fate of the Universe. For instance, we know that if (in the $\Lambda$CDM model) the cosmological constant had become dominant earlier than it really did then structures would not have formed. Insights into these matters may put us on the track of a solution of puzzles like the coincidence problem in the sense that it would be good to find some physical mechanism associating the beginning of the dark ages with the onset of non-linear structures.

Our first task will go back to the previous discussion in Sec. \ref{section:fqmodels} in which an effective dark energy density $\rho_{\rm eff}$ and corresponding equation of state $w_{\rm eff}$ were defined. The explicit expressions for the two cases considered are formidable and they really do not add readily usable information to the discussion. For this reason we just rely on graphical representations of $w_{\rm eff}$ drawn from its best fit and errors (see Table  \ref{Cosmographicvalues}). The plots we thus obtain confirm our earlier conclusions for all redshifts that in the DGP case (with $\alpha<0$) we are to expect a non-phantom behavior, whereas in the DGPish case  (with $\gamma<0$) we have right the opposite, a purely phantom evolution.

Next we will determine numerically the
redshift value signaling the beginning of dark energy domination: $z_{eq}$. To this end we rest on our earlier definition of $\rho_{\rm eff}$ and we then define 
$z_{eq}$ implicitly as
\bea
\rho_{\rm eff}(z_{eq})=\rho_m(z_{eq}).
\eea As discussed on \cite{Melchiorri:2007in}, evidences of considerable dark energy proportions at $z\sim 1$ or larger   would favor some scalar field based models, as for those redshifts the contribution of a cosmological constant to the matter-energy budget would be negligible.  On the contrary, and according to  \cite{Melchiorri:2007in} again, a beginning of dark energy domination occurring at $z\sim 0.2$ or lower seems to indicate phantom dark energy. These assertions depend of course on the value of $\Omega_m$, and, in general, the discussion may branch out into the specifics of the effects of the various parameters entering our dark energy description. 

Likewise, it might be interesting to learn not only when does dark energy domination begin, but also when do its effects become manifest, that is, which is the redshift value for which  acceleration begins ($z_{acc}$). This occurs when the so called deceleration parameter $q$ vanishes ($q(z_{acc})=0$) \footnote{For $\Lambda$CDM and provided that the contribution from radiation is considered to be negligible, the pertinent expressions for the beginning of the acceleration $q(z_{acc})=0$ (which implies $z_{acc}=-1+(2\Omega_\Lambda/\Omega_m)^{1/3}$)  and  the condition of the onset of dark energy domination $\Omega_\Lambda=(1+z_{eq})^3 \Omega_m$ can be combined  to conclude $z_{acc}=-1+2^{1/3}(1+z_{eq})$.}. Just recall for calculation purposes that
\bea
q(t)=-\displaystyle\frac{a\ddot{a}}{\dot{a}^2} \;\rightarrow \; &q(z)=-1+(1+z)\displaystyle\frac{E'(z)}{E(z)}
\eea

Again, the explicit expressions are too lengthy and convoluted to be worth presenting here, and once more we report our findings numerically and graphically. 

It is customary to evaluate the former quantity at $z=0$ along with equivalent values of the other parameters in the (standard) cosmographic cascade, here we just add our findings on the next one, the jerk:
\bea
j(t)=\displaystyle\frac{\dot{\ddot{a}}a^2}{\dot{a}^3} \;\rightarrow \;&j(z)=(1+z)^2\displaystyle\frac{E''(z)}{E(z)}+q^2(z)
\eea

The normalization condition can be used to eliminate from expressions the parameters that are less convenient/important for our discussion. In the case of the effect of the $f(Q)$ correction on the deceleration factor and the jerk of each model we choose to eliminate $\Omega_{\Lambda}$. More specifically at $z=0$:

\begin{widetext}
\begin{align}
  &q_0=q_0^{\Lambda{\rm CDM}}
      -    \left\{\begin{array}{l}        \displaystyle  \frac{3}{2} (3 \Omega_m + 4 \Omega_r) \Omega_Q+{\cal O}(Q^2)  \qquad \hbox{for the DGP model}
       \\        \\
      \displaystyle  \frac{1}{4}\left(3\Omega_m+4\Omega_r\right)\Omega_Q^2+{\cal O}(Q^3)\qquad \hbox{for the DGPish model}
   \end{array}\right. 
    \\\nonumber
    \\\nonumber
    &\hbox{where}\;\;\; q_0^{\Lambda{\rm CDM}}=-1+\frac{1}{2}\left(3\Omega_m+4\Omega_r\right)\\ \nonumber \\
      &j=j_0^{\Lambda{\rm CDM}}- 
    \left\{\begin{array}{l}
        \displaystyle  \frac{1}{4}\left[24\Omega_r+(3\Omega_m+4\Omega_r)^2\right]\Omega_Q+{\cal O}(Q^2)  \qquad \hbox{for the DGP model}
        \\
        \\
      \displaystyle  \frac{1}{2}\left[2\Omega_r-(3\Omega_m+4\Omega_r)^2\right]\Omega_Q^2+{\cal O}(Q^3)\qquad \hbox{for the DGPish model}
    \end{array}\right. 
    \\\nonumber
    \\\nonumber
    &\hbox{where}\;\;\; j_0^{\Lambda{\rm CDM}}=1+2\Omega_r
\end{align}
\end{widetext}

Using the latter expression  we can compare analytically  the deceleration factor for both modified gravity models  and a $\Lambda$CDM one with the same values of the parameters $\Omega_m$ and  $\Omega_r$.  As it can be noticed, for the DGP model with negative $\Omega_Q$, the deceleration parameter gets bigger, i.e.  $q_0>q_0^{\Lambda CDM}$ and consequently the acceleration is smaller. However, for the DGPish model the sign of $\Omega_Q$ does not matter any more, the deceleration parameter will always be  smaller than in $\Lambda$CDM which gives comparably a more pronounced acceleration.


All the previous results are summarized in the next tables and contours:

\begin{widetext}

{\renewcommand{\tabcolsep}{1.mm}
{\renewcommand{\arraystretch}{1.5}
\begin{table*}[htbp]
\caption{Best fits and errors of the cosmographic parameters.}
\begin{tabular}{c ccccc}
\hline\hline
    & \multicolumn{1}{c}{$q_0$} &\multicolumn{1}{c}{$j_0$} 
    &\multicolumn{1}{c}{$w_{\rm eff}$}
    &\multicolumn{1}{c}{$z_{ac}$}
    &\multicolumn{1}{c}{$z_{eq}$}
    \\ \hline
  $\Lambda$CDM &
 $-0.515_{-0.007}^{+0.007}$ &
$1.000186_{-2*10^{-6}}^{+2*10^{-6}}$ & 
  $-1$ &  $0.612_{-0.012}^{+0.012}$ & $0.280_{-0.009}^{+0.009}$
   \\ \cline{1-6}
\multicolumn{1}{l}{}           $DGP$   &
$-0.499_{-0.017}^{+0.018}$ &
$0.994_{-0.007}^{+0.006}$ &
$-0.989_{-0.011}^{+0.012}$ & $0.593_{-0.022}^{+0.022}$ & $0.278_{-0.010}^{+0.010}$
   \\ \cline{1-6}
\multicolumn{1}{l}{}         $DGPish$   &
$-0.523_{-0.015}^{+0.010}$ &
$1.009_{-0.008}^{+0.029}$ &
$-1.004_{-0.012}^{+0.004}$ &  $0.614_{-0.012}^{+0.012}$ & $0.279_{-0.009}^{+0.010}$
\\ \hline
\end{tabular}
\label{Cosmographicvalues}
\end{table*}}}

\begin{figure*}[htb]
    \includegraphics[width=0.4\linewidth]{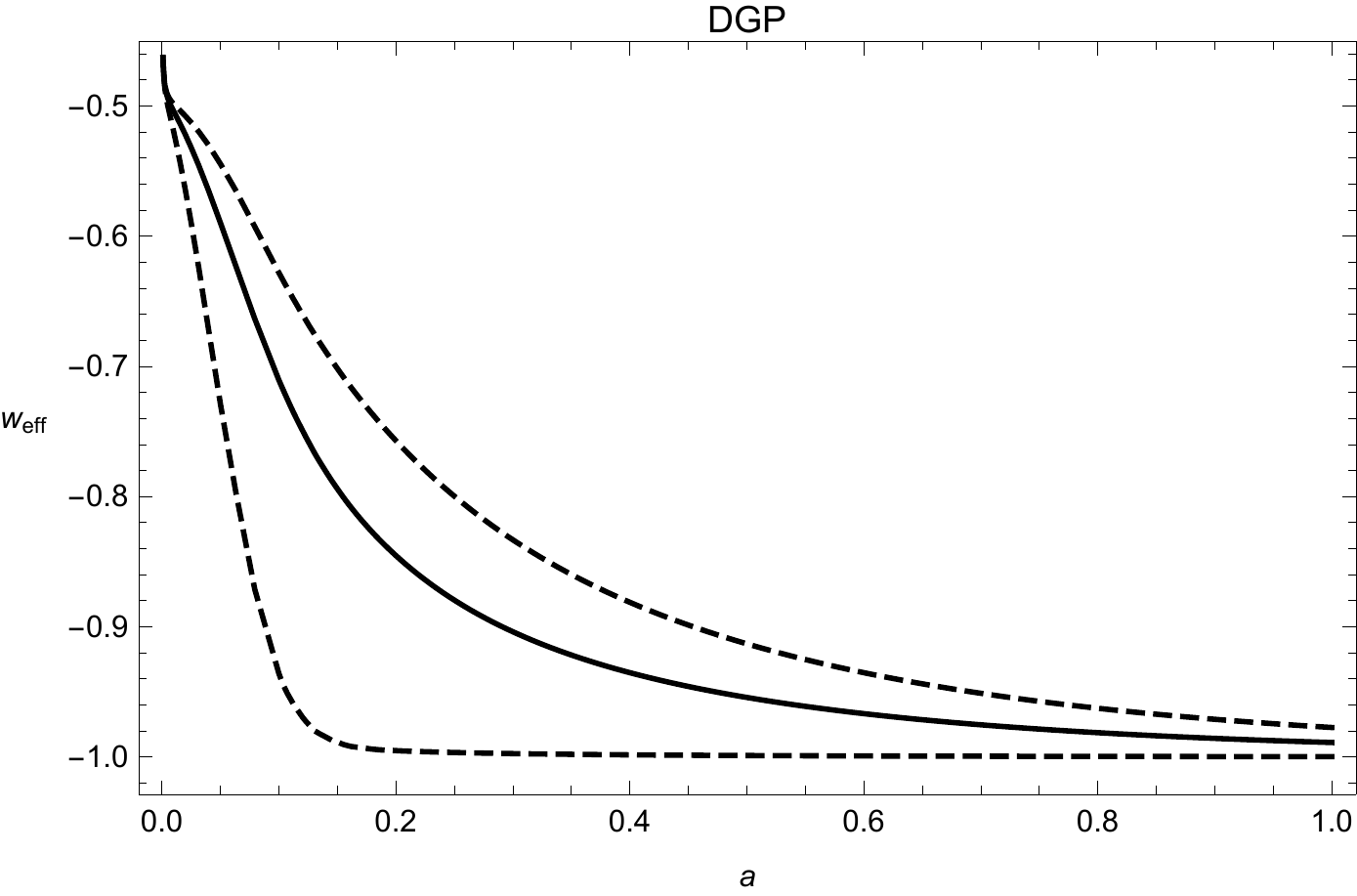}~~~
    \includegraphics[width=0.4\linewidth]{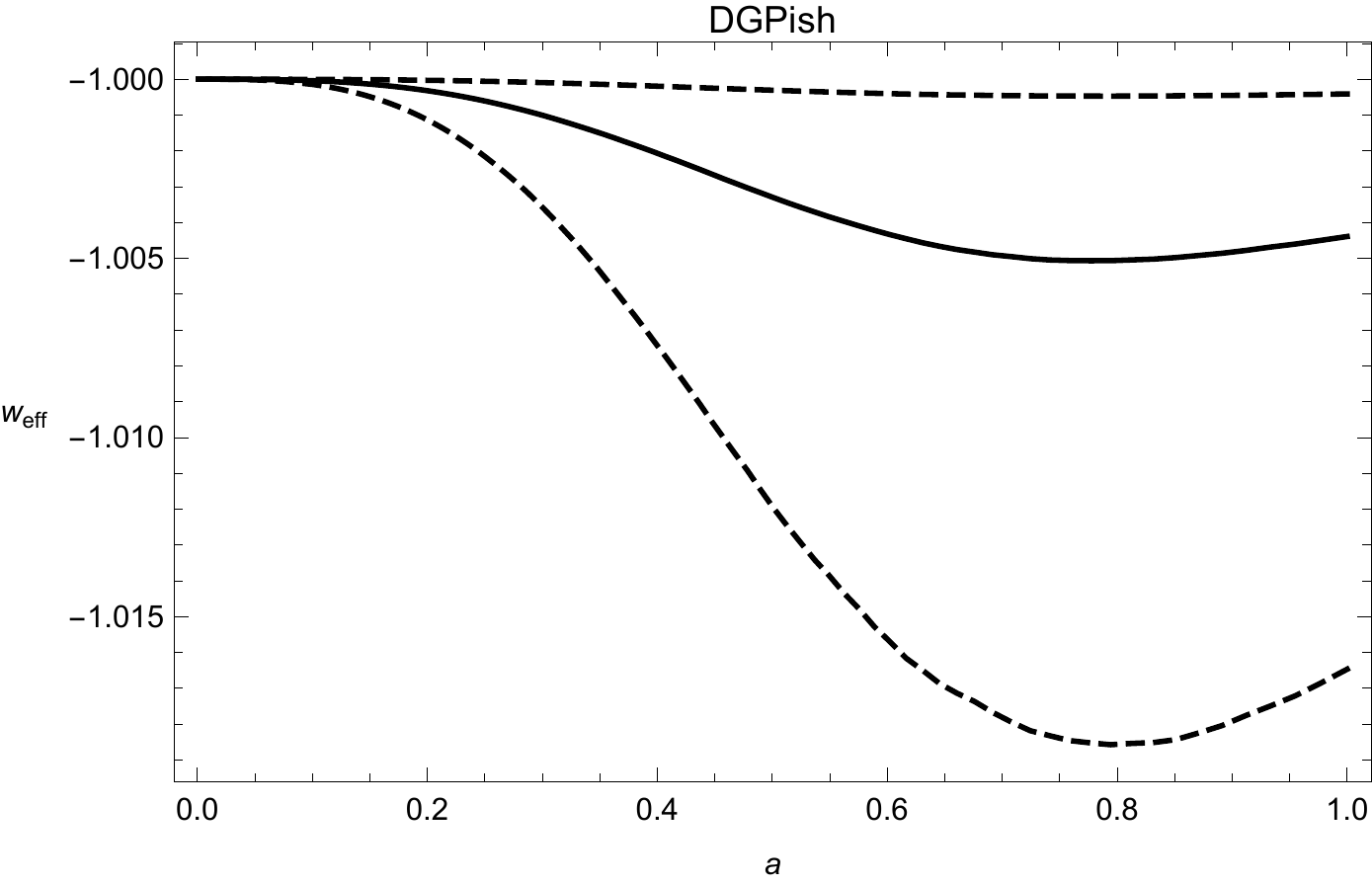}~~~\\
    \caption{Evolution of $w_{\rm eff}$ as a function of the scale factor $a$ for $DGP$ (left panel) and $DGPish$ (right panel) model. The solid line is the value as drawn from the best fit, whilst the dashed lines mark the boundaries of the confidence interval.}
    \label{weff}
\end{figure*}

\end{widetext}

\begin{figure}[htb]
    \includegraphics[width=0.9\linewidth]{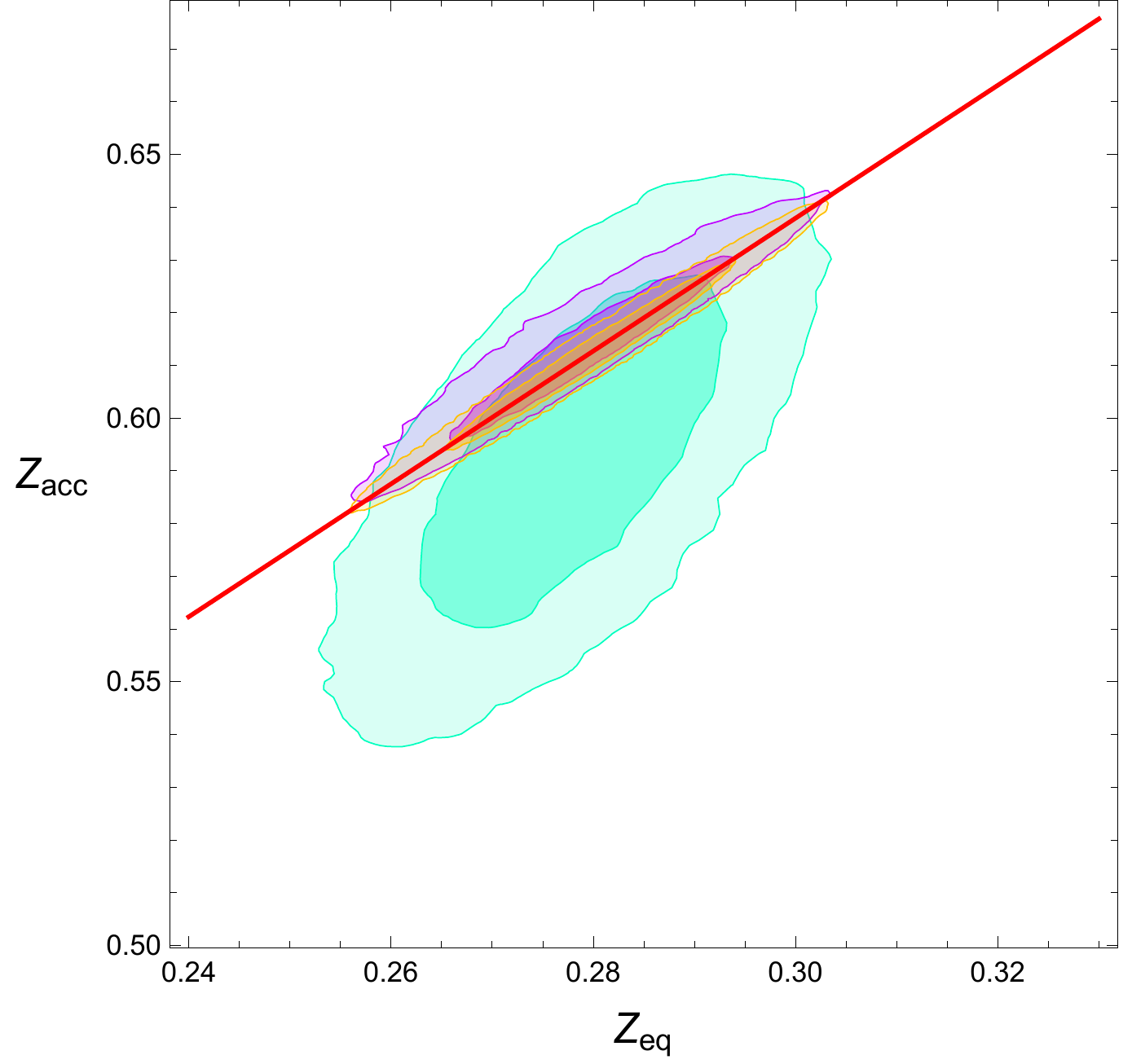}~~~
    \caption{Confidence contours for the $z_{eq}$ and $z_{acc}$ redshift values of the different models as drawn from the MCMC results: $\Lambda$CDM (orange), DGP (blue) and DGPish (purple). The red line indicates the theoretical relation for the  $\Lambda$CDM case.}
    \label{Z}
\end{figure}

\section{Conclusions}\label{Conclusions}

Along this work we have analyzed two cosmological models stemming from  modified gravity proposals in the $f(Q)$ arena. One of them has an evolution which is identical to that of the DGP models, although the different Lagrangians they are derived from point out in the direction of differences at the level of perturbations, which are beyond our scope. The second model bears a considerable resemblance to the former, and for that reason we have branded it as DGPish. We offer some details regarding the prescription to generate new $f(Q)$ cosmological models from a few simple assumptions with some physical guidance.

Through MCMC best fits of their free parameters and the use of  selection criteria in a analogous way to our previous work \cite{Ayuso:2020dcu} we are able to discern whether these models can be viewed as contenders to the $\Lambda$CDM evolution. 

By encoding the new features of the models into a singles parameter dubbed as $\Omega_Q$  the comparison can be done efficiently. Results indicate a statistical preference of negligible values of both $f(Q)$ models, so much so for individual data as for data set combined. The only case in which $\Omega_Q$  seems to find most confortable a little far away from zero is when the analysis is performed with BAO data for the DGP model. All in all, the possible new phenomenology associated with the $f(Q)$ features finds room between possible evolutions for our universe but does not seem to be preferred, at least along the two routes we have explored. 
This fact gets reinforced by values we obtain for the Bayesian evidence, which according to Jeffreys' scale, tell us again these models are not preferred over the $\Lambda$CDM one. 

However the mildly better agreement on $\Omega_m$, $\Omega_b$ and $h$ values between the  DGPish and the $\Lambda$CDM suggest that perhaps a generalization with extra parameters could offer an even better agreement while retaining some modified gravity charc. 

In addition, we have performed a cosmographic study with similar results  which reflect the striking similarity of the best fits between $\Lambda$CDM and our $f(Q)$ models once more, with similar or bigger errors due since their complexity, which penalizes error propagation.

Therefore, at the end of the day, both models are so good as $\Lambda$CDM at the background level but only when they fall upon $\Lambda$CDM, which is obvious but does not turn on the new phenomenology or with other words, this new phenomenology is no necessary. Hence, these both $f(Q)$ models cannot be considered as new better models, at least as far as the background considerations.

\section*{Acknowledgments}
We are grateful to Vincenzo Salzano for enlightening discussions.   IA was funded by Fundação para a Ciência e a Tecnologia (FCT) grant PD/BD/114435/2016 under the IDPASC PhD Program. RL was supported by the Spanish Ministry of Science and Innovation through research projects  FIS2017-85076-P (comprising FEDER funds), and also by the Basque Govern\-ment and Generalitat Valenciana through research projects  GIC17/116-IT956-16 and  PROMETEO/2020/079 respectively. IA and JPM acknowledge the research grants No. UID/FIS/04434/2020, No. PTDC/FIS-OUT/29048/2017, and No. CERN/FIS-PAR/0037/2019.





%

\end{document}